\documentclass[prl, aps, longbibliography, twocolumn, amsmath, amssymb, superscriptaddress,showpacs]{revtex4-1} 
\usepackage{graphicx, hyperref, color} 
 
\begin{document}

\title{Level Attraction Due to Dissipative Magnon-Photon Coupling}

\author{M. Harder}
\affiliation{Department of Physics and Astronomy, University of Manitoba, Winnipeg, Canada R3T 2N2}

\author{Y. Yang}
\affiliation{Department of Physics and Astronomy, University of Manitoba, Winnipeg, Canada R3T 2N2}
\affiliation{State Key Laboratory of Infrared Physics, Chinese Academy of Sciences, Shanghai 200083, People's Republic of China}

\author{B. M. Yao}
\affiliation{State Key Laboratory of Infrared Physics, Chinese Academy of Sciences, Shanghai 200083, People's Republic of China}

\author{C. H. Yu}
\affiliation{Department of Physics and Astronomy, University of Manitoba, Winnipeg, Canada R3T 2N2}
\affiliation{Jiangsu Key Laboratory of ASIC Design, Nantong University, Nantong 226019, China}

\author{J. W. Rao}
\affiliation{Department of Physics and Astronomy, University of Manitoba, Winnipeg, Canada R3T 2N2}

\author{Y. S. Gui}
\affiliation{Department of Physics and Astronomy, University of Manitoba, Winnipeg, Canada R3T 2N2}

\author{R. L. Stamps}
\affiliation{Department of Physics and Astronomy, University of Manitoba, Winnipeg, Canada R3T 2N2}

\author{C. -M. Hu}
\email{Corresponding author. Email: hu@physics.umanitoba.ca}
\affiliation{Department of Physics and Astronomy, University of Manitoba, Winnipeg, Canada R3T 2N2}

\date{\today}

\begin{abstract} 

We report dissipative magnon-photon coupling caused by cavity Lenz effect, where the magnons in a magnet induce a rf current in the cavity, leading to a cavity back action that impedes the magnetization dynamics. This effect is revealed in our experiment as level attraction with a coalescence of hybridized magnon-photon modes, which is distinctly different from level repulsion with mode anticrossing caused by coherent magnon-photon coupling. We develop a method to control the interpolation of coherent and dissipative magnon-photon coupling, and observe a matching condition where the two effects cancel. Our work sheds light on the so-far hidden side of magnon-photon coupling, opening a new avenue for controlling and utilizing light-matter interactions.        

\end{abstract}

\maketitle

The recent discovery of coherent magnon-photon coupling has spawned the rapid development of cavity-spintronic technologies \cite{Soykal2010,Huebl2013a,Zhang2014,Tabuchi2014a,Goryachev2014,Bai2015,Zhang2015g,Haigh2015,Hu2015,MaierFlaig2016,Osada2016,Zhang2016,Tabuchi2017,Yao2017,Bai2017, Wang2018}, lying at the crossroads of quantum information and spintronics.  On one hand, it has enabled a gradient memory architecture \cite{Zhang2015g}, single magnon detection \cite{Tabuchi2017}, and dressing of magnon dressed states \cite{Yao2017}. All of these could be exploited for magnon based quantum information applications.  On the other hand it led to the generation of spin current from the cavity magnon polariton \cite{Bai2015, MaierFlaig2016}, which has created novel spintronic applications at room temperature, such as non-local spin current manipulation \cite{Bai2017}.  

The physics of coherent magnon-photon coupling can be understood, either quantum mechanically as spin-photon entanglement \cite{Soykal2010, Huebl2013a}, or classically as mutual coupling between electro and magnetization dynamics \cite{Bai2015}. Consistent with the correspondence principle, both pictures equally explain the characteristic coupling features such as level repulsion, damping exchange, and Rabi oscillations. Despite its broad impact \cite{Soykal2010,Huebl2013a,Zhang2014,Tabuchi2014a,Goryachev2014,Bai2015,Zhang2015g,Haigh2015,Hu2015,MaierFlaig2016,Osada2016,Zhang2016,Tabuchi2017,Yao2017,Bai2017, Wang2018}, coherent coupling is just the tip of the iceberg of magnon-photon hybridization. Here, we reveal a dissipative magnon-photon coupling, whose distinct coupling features have so-far been hidden. 

Let us first explain the central idea by using a textbook example and a \textit{gedanken} experiment.

As schematically shown in Fig.~1(a), the descent of a magnet dropping inside a metallic pipe is impeded by the induced current, which gives rise to a magnetic back action that opposes the change in original magnetic flux. That's Lenz's law. In a gedanken experiment if we replace the moving magnet with the precessional magnetization, and the pipe with a microwave cavity as shown in Fig.~1(b), then the back action of the induced current shall impede the magnetization dynamics, so that the magnons shall be coupled with the induced cavity current via the damping-like Lenz effect. This we refer to as dissipative magnon-photon coupling, in contrast to the coherent coupling effect where the cavity current drives the magnetization dynamics \cite{Bai2015}.   

\begin{figure}[!t]
\includegraphics[width = 8.7 cm]{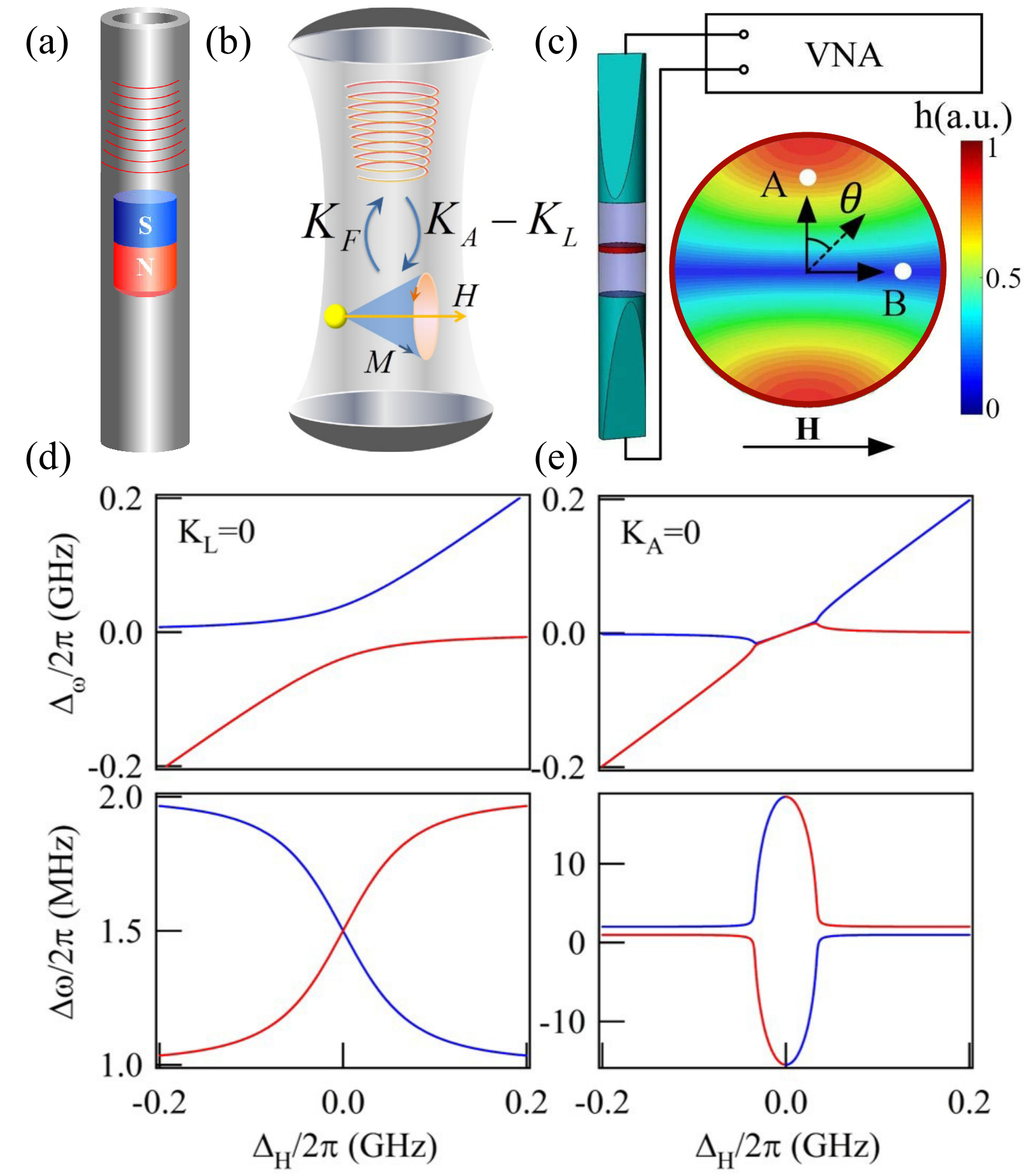} 
\caption{(a) The Lenz effect.  The motion of a magnet falling down a conducting pipe is impeded by the induced magnetic field.  (b) Magnon-photon coupling mechanisms including the cavity Lenz effect that impedes the magnetization dynamics.  (c) Experimental setup, with a VNA measuring the microwave transmission through a waveguide loaded with a YIG sphere.  The simulated $h$ field amplitude for the TE$_{11}$ mode at the middle plane of the empty waveguide.  A and B denote the $h$ antinode and node positions for loading YIG sphere.  (d) and (e) The hybridized mode frequency ($\Delta_\omega \equiv \omega - \omega_c$) and line width ($\Delta\omega$) are plotted as a function of the field detuning [$\Delta_\text{H} \equiv \omega_r(H)- \omega_{c}$], for level repulsion and attraction predicted by solving Eq. \eqref{coupling} at $K_L$ = 0 and $K_A$ = 0, respectively. Sample parameters are given in the paper.} 
\label{fig:fig1}
\end{figure}

Such a cavity Lenz effect has never been observed \cite{Note}. Instead, recent experiments \cite{Huebl2013a,Zhang2014,Tabuchi2014a,Goryachev2014,Bai2015,Zhang2015g,Haigh2015,Hu2015,MaierFlaig2016,Osada2016,Zhang2016,Tabuchi2017,Yao2017,Bai2017, Wang2018} with a magnet in a cavity measure the combined Amp$\grave{e}$re and Faraday effect, which lead to coherent magnon-photon coupling as we've explained \cite{Bai2015}. Now, considering the cavity Lenz effect, and using a similar approach as in Ref. \cite{Bai2015} by combining the RLC and LLG equations for the cavity current $j$ and dynamic magnetization $m$, respectively, we get the following eigenvalue equations \cite{Suppl}:
\begin{equation}
\begin{pmatrix} \omega^2-\omega_c^2+i2\beta\omega_c\omega & i\omega^2K_F\\-i\omega_m(K_A-K_L) & \omega-\omega_r+i\alpha\omega \end{pmatrix} \begin{pmatrix} j \\ m \end{pmatrix} = 0,
\label{coupling}
\end{equation}
where $\omega_m$ = $\gamma M_0$ is proportional to the saturation magnetization $M_0$, and $\gamma$ is the gyromagnetic ratio. 

Equation (1) describes the coupling between the magnon mode at $\omega_r$ and the cavity mode at  $\omega_c$, which have an intrinsic damping rate of $\alpha$ and $\beta$, respectively. The $K_F$-term stems from Faraday's law \cite{Bai2015}. It describes the effect of the dynamic magnetization $m$ on the rf current $j$.  The $K_A$-term comes from Amp$\grave{e}$re's law. It shows that the current $j$ produces a rf magnetic field, which drives the magnetization via a field torque \cite{Bai2015}. Cavity Lenz effect appears in the $K_L$-term, which has the opposite sign of the $K_A$-term, since the back action from the induced rf current impedes the magnetization dynamics, instead of driving it \cite{Suppl}.    

Ignoring the cavity Lenz effect by setting $K_L$ = 0, the eigen solution \cite{Suppl} of Eq.~(\ref{coupling}) reproduces the coherent magnon-photon coupling \cite{Bai2015}, characterized by level repulsion and damping exchange, as shown in Fig.~1(d). The coupling strength $g_A = \sqrt{\omega_c\omega_m K_F K_A/2}$ determines the Rabi splitting. On the other hand, by setting $K_A$ = 0, the eigen solution of Eq.~(\ref{coupling}) reveals the dissipative magnon-photon coupling characterized by level attraction \cite{Glauber1985} and damping repulsion, as shown in Fig.~1(e). The coupling strength is given by $g_L =\sqrt{\omega_c\omega_m K_F K_L/2}$, which determines the damping splitting \cite{Suppl}. With both Amp$\grave{e}$re and Lenz terms, level repulsion and attraction appear when $K_A - K_L >$ 0 and $K_A - K_L <$ 0, respectively. These two regimes of magnon-photon coupling are separated by a matching condition at $K_A - K_L$ = 0, where the magnons and photons appear as decoupled.   

It is clear therefore, revealing the cavity Lenz effect requires suppressing the torque related to the Amp$\grave{e}$re term $K_A$, based on which we design our experiment.  

Our setup is shown in Fig. \ref{fig:fig1} (c).  We connect a Ku-band circular waveguide with circular-rectangular transitions to form a 1D Fabry-Perot-like cavity.  Such a special cavity offers an excellent mode profile controllability as we demonstrated in Ref. \cite{Yao2015}.  The inner diameter of the circular waveguide is 16.1 mm. Magnon-photon coupling is studied by placing a 1-mm diameter YIG sphere ($\mu_0M_0 = 0.178$ T) approximately 3 mm from the inner edge at the middle plane of the waveguide.  The magnon mode with $\alpha = 7.60 \times 10^{-5}$ follows the dispersion $\omega_r(H) = \gamma \left(H + H_A\right)$, where $\gamma = 2 \pi \times 27 ~\mu_0$GHz/T, $\mu_0H_A = 6$ mT is the magneto-crystalline anisotropy field and $H$ is the static bias magnetic field applied perpendicular to the cavity axis.

To suppress the Amp$\grave{e}$re term we use the TE$_{11}$ mode of this cavity at $\omega_c/2\pi = 13.205$ GHz. The intrinsic damping rate of this mode is $\beta = 1.50 \times 10^{-4}$ (Q = 3300). When loaded with the YIG sphere and connected for measurements, the damping rate increases to $\beta_\text{L} = 8.49 \times 10^{-3}$. Without inserting the YIG sphere, as calulated by Computer Simulation Technology (CST) in Fig. \ref{fig:fig1} (c), the TE$_{11}$ mode at the middle plane of the waveguide has a maximum (position A) in the microwave magnetic field amplitude, $h$, where the Amp$\grave{e}$re term $K_A$ is large. This is where coherent magnon-photon coupling is usually measured \cite{Huebl2013a,Zhang2014,Tabuchi2014a,Goryachev2014,Bai2015,Zhang2015g,Haigh2015,Hu2015,MaierFlaig2016,Osada2016,Zhang2016,Tabuchi2017,Yao2017,Bai2017, Wang2018}. The TE$_{11}$ mode also has an $h$-node (position B) with small $K_A$. This is the position where we set the YIG sphere to explore the level attraction induced by the cavity Lenz effect. 

\begin{figure}[!t]
\includegraphics[width = 8.7 cm]{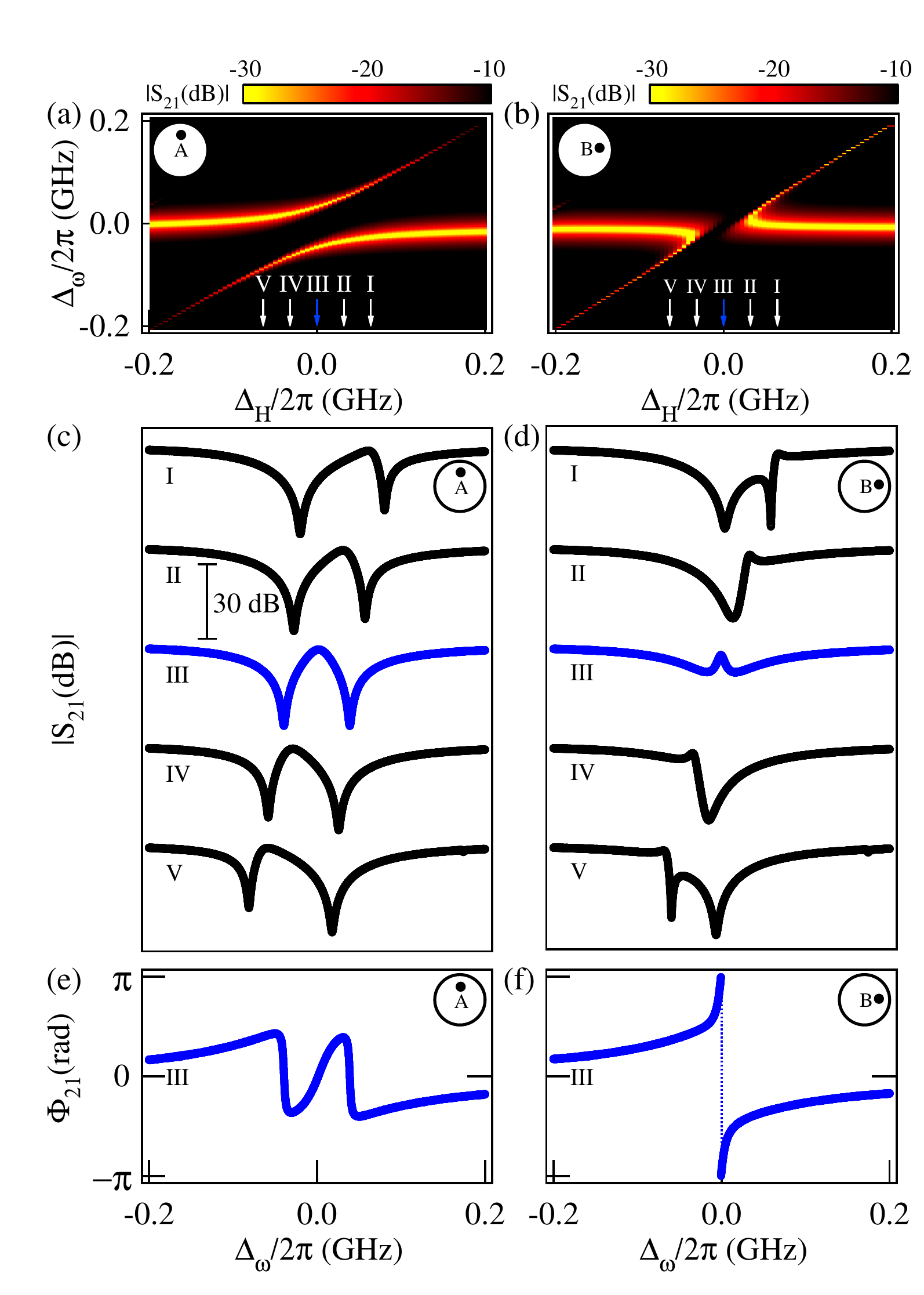} 
\caption{Experimental transmission spectra mappings as a function of $\Delta_\omega$ and $\Delta_H$ for (a) level repulsion and (b) level attraction.  The YIG sphere has been placed at positions A and B to observe level repulsion and attraction respectively.  Individual transmission spectra at different field detunings are shown for (c) level repulsion and (d) level attraction.  (e) The phase behaviour of level repulsion and (f) level attraction measured at $\Delta_\text{H} = 0$.}
\label{fig:fig2}
\end{figure}

Using a vector network analyzer (VNA) we measure the microwave transmission $S_{21}$ of the magnon-photon system.  With the YIG located at the $h$ antinode, position A in Fig. \ref{fig:fig1}(c), we observe conventional level repulsion of the hybridized modes.  Here, a mapping of $|S_{21}|$ is plotted in Fig. \ref{fig:fig2}(a) as a function of the frequency and field detuning, $\Delta_\omega = \omega - \omega_c$ and $\Delta_\text{H} = \omega_r(H)- \omega_{c}$.  From the Rabi splitting \cite{Harder2016b} we obtain a coupling strength of $g_A/2\pi = 39$ MHz.  Strikingly, when the YIG sample is placed at the $h$-node position B, we no longer observe level repulsion, but instead discover level attraction as shown in Fig. \ref{fig:fig2} (b). We measure a coupling strength of $g_L/2\pi = 17$ MHz. 

The spectral contrast between level repulsion and level attraction is shown in Fig. \ref{fig:fig2}(c) and (d).  Here we plot $S_{21}$ as a function of $\omega$ for several values of the field detuning $\Delta_\text{H}$.  The level repulsion behaviour is shown in panel (c).  At $\Delta_\text{H}$ = 0 two equal amplitude resonances are observed and as $|\Delta_\text{H}|$ increases these two modes move apart with one mode slowly decreasing in amplitude, consistent with the observed modes anticrossing.  In contrast, as shown in panel (d) for level attraction, a sharp peak is superimposed on a broad resonance, and at $\Delta_\text{H}$ = 0 both appear at the same frequency $\omega = \omega_r = \omega_{c}$. An even more striking contrast can be observed by examining the phase $\phi_{21}$ for $S_{21}$ measured at $\Delta_\text{H}$ = 0. In the case of level repulsion as shown in panel (e), we observe two $\pi$-phase shifts at each of the distinct hybridized modes \cite{Harder2016}.  However in the case of level attraction shown in panel (f), we observe a single $2\pi$-phase jump, confirming the fact that two hybridized modes have coalesced at $\omega = \omega_r = \omega_{c}$.  

Having verified the presence of both level repulsion and level attraction, we now turn to the other key prediction of Eq. (1):  an experimentally accessible competition between the two competing magnon-photon coupling mechanisms, which leads to a matching condition separating the two coupling regimes. This we demonstrate by using a specially designed waveguide insert, which allows us to systematically tune the angular position $\theta$ of the YIG within a $1^\circ$ precision, as shown in Fig. 1(c).  This insert increases the length of the circular waveguide by 8.5 mm, resulting in a 3.6 \% red shift of the cavity frequency ($\omega_c/2\pi = 12.725$ GHz) and a slightly changed cavity damping of $\beta_\text{L} = 6.54 \times 10^{-3}$.  

\begin{figure}[!t]
\includegraphics[width = 9.0 cm]{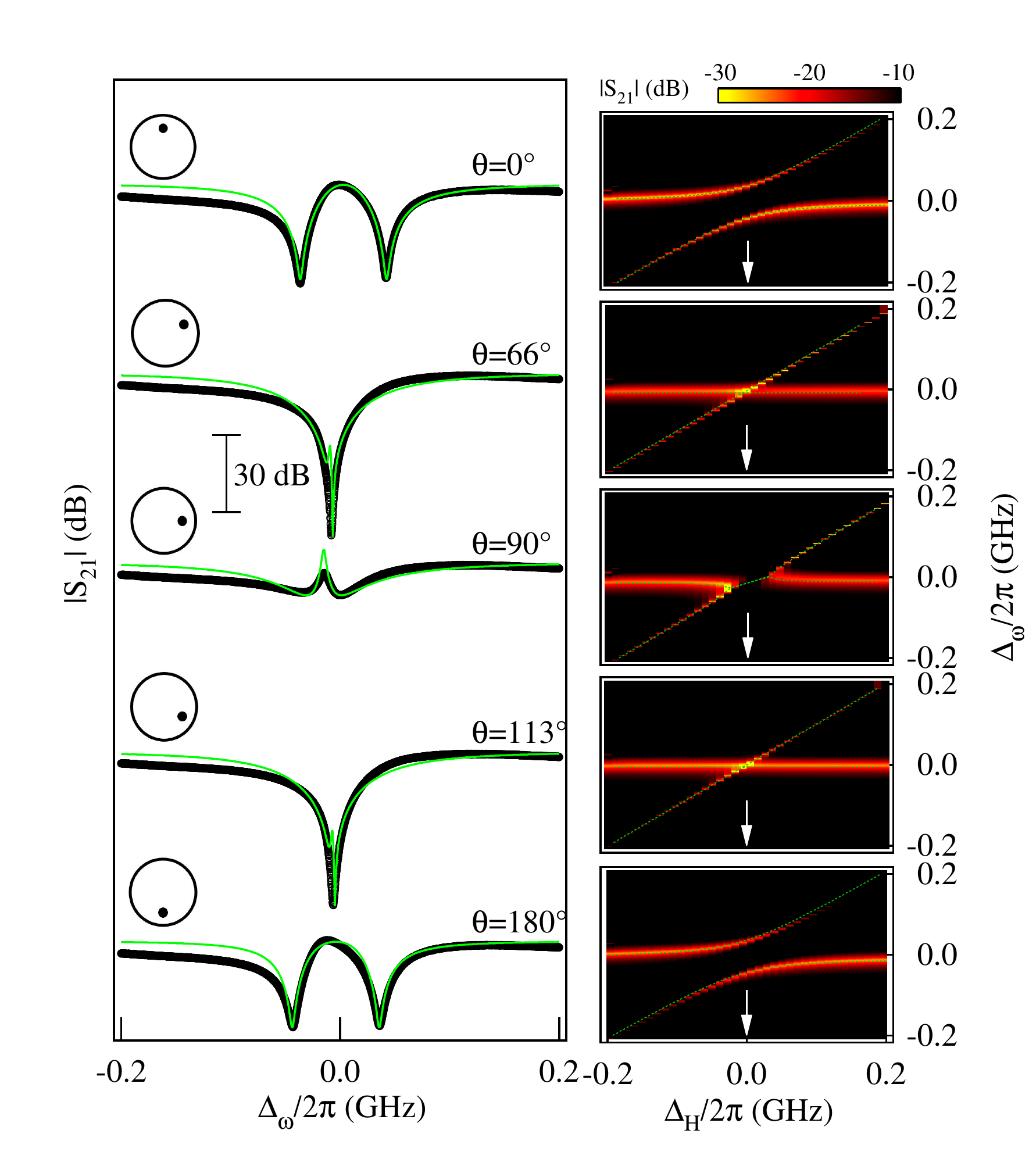} 
\caption{Transmission spectra at various angular positions $\theta$, demonstrating the change from level repulsion to level attraction.  Thick black curves are experimental data and thin green curves are calculations.  The corresponding $\Delta_\omega - \Delta_\text{H}$ transmission mappings are shown on the right.} 
\label{fig:fig3}
\end{figure}

The transmission spectra at different $\theta$ are shown in Fig. \ref{fig:fig3}, all measured at $\Delta_\text{H} = 0$. The corresponding full mappings are plotted in the right panel.  At both $\theta = 0$ and 180$^\circ$, level repulsion is observed. At $\theta = 90^\circ$, level attraction is observed. As we change $\theta$, two matching conditions are found near $\theta \simeq$ 65$^\circ$ and 115$^\circ$, where the coupling strength is very small, resulting in level crossing as shown by the $|S_{21}|$ mapping. Overall, as will be analyzed in detail below, we find that within the range $0^\circ < \theta < 180^\circ$, level attraction appears in the region of $65^\circ < \theta < 115^\circ$, which is separated from the other regions of level repulsion by the matching conditions. 

Thus far, based on Eq. (1), we have explained the physical origin of coherent and dissipative magnon-photon coupling on the equal footing of classical electrodynamics. And we have shown that the two competing coupling mechanisms lead to two distinct types of mode hybridization: level repulsion and level attraction. Note that  Eq. (1) stems from two linearly coupled harmonic oscillators \cite{Harder2016b}. In light of the Ehrenfest theorem, we further develop a quantum picture by constructing the Hamiltonian: 
\begin{equation}
H = \hbar \omega_c a^\dagger a + \hbar \omega_r b^\dagger b + \hbar g \left(a^\dagger b + e^{i\Phi} b^\dagger a\right), \label{eq:hamiltonian}
\end{equation}
where $a^\dagger$ ($a$) and $b^\dagger$ ($b$) are the creation (annihilation) operators for cavity-photons and ferromagnetic magnons respectively. Here, the coupling phase $\Phi$ describes the competing effect of two forms of magnon-photon coupling, and $g$ is the net coupling strength. 

Equation (2)  places the physics of magnon-photon coupling in a broad context: when $\Phi = 0$ the coupling term describes the ubiquitous coherent coupling that forms quasi-particles such as polaritons \cite{Huebl2013a, Harder2016b} and polarons \cite{Holanda2018}; With $\Phi = \pi$ the coupling term resembles the dissipative coupling \cite{Metelmann2014}  recently realized in cavity optomechanical systems \cite{Gloppe2014,Xu2016}. The eigen frequencies of Eq. (2) are
\begin{equation}
\omega_\pm = \frac{1}{2}\left[\omega_c + \omega_r \pm \sqrt{\left(\omega_r - \omega_c\right)^2 + 4 e^{i \Phi} g^2}\right]. \label{eq:dispersion}
\end{equation}
At $\Phi$ = 0 and $\pi$, Eq. (3) agrees very well with the solutions of Eq. (1) plotted in Fig. 1(d) and (e), respectively, by setting $g^2 = \omega_c\omega_m |K_F \left(K_A- K_L\right)|/2$.

\begin{figure}[!t]
\includegraphics[width = 8.7 cm]{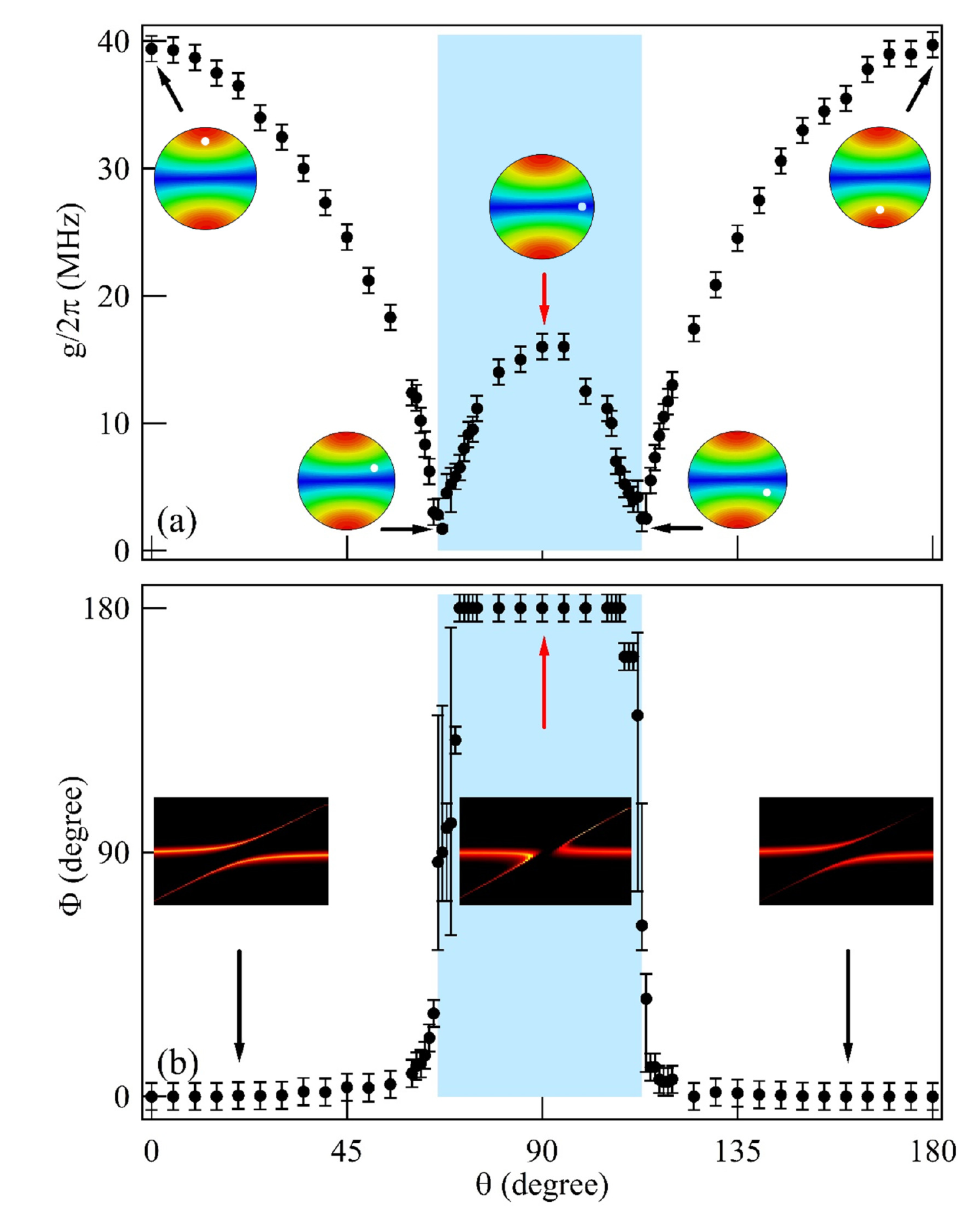} 
\caption{Systematic control of the coupling strength and coupling phase in our set-up.  (a) The net coupling strength, $g$, measured at different YIG positions, $\theta$, as indicated by the insets.  (b) The coupling phase, $\Phi$, measured at different YIG position, $\theta$, reveals two distinct regions for level repulsion and attraction where $\Phi = 0$ and $\pi$ respectively.} 
\label{fig:fig4}
\end{figure}

As shown in Fig. 3, by using Eq. (3) to fit the measured dispersions, and also by using $S_{21}(\omega)$ calculated  from Eq. (2) to fit the measured transmission spectra  \cite{Suppl, Harder2016b}, we determine the coupling strength, $g$, and the coupling phase, $\Phi$, both as a function of the angular position $\theta$ for the YIG sphere.  The results are summarized in Fig. \ref{fig:fig4} with insets in panel (a) indicating the YIG position in the waveguide field.  

Clearly, three key features of magnon-photon coupling stand out: (i)  Two distinct coupling regions: the level attraction region (shadowed) characterized by $\Phi = \pi$, appears in the region of $65^\circ < \theta < 115^\circ$, which is separated from the level repulsion region characterized by $\Phi = 0$. (ii) Within each region, the net coupling strength $g$ varies at different angular positions. (iii) The two regions are sharply separated by the matching condition appearing at $\theta \simeq$ 65$^\circ$ and 115$^\circ$, where $g$ diminishes and the measured $\Phi$ is uncertain. All these features are consistent with the results discussed based on the classical picture, showing clearly that two competing magnon-photon coupling effects coexist at general experimental conditions. 

Thus, our study reveals the cavity Lenz effect that leads to dissipative magnon-photon coupling. Distinct features, including level attraction with coalescence of the hybridized dispersions, are observed. By developing consistent models built on both classical and quantum mechanical formalisms, we establish a comprehensive picture for understanding magnon-photon coupling, which is currently of great interest. Revealing such a hidden dissipative nature enables a new way for controlling magnon-photon hybridization, which we demonstrate by tuning the interpolation between coherent and dissipative magnon-photon coupling. Our results show that even in the conventional level repulsion regime \cite{Soykal2010,Huebl2013a,Zhang2014,Tabuchi2014a,Goryachev2014,Bai2015,Zhang2015g,Haigh2015,Hu2015,MaierFlaig2016,Osada2016,Zhang2016,Tabuchi2017,Yao2017,Bai2017, Wang2018}, dissipative magnon-photon coupling competes with coherent coupling, which leads to a reduced net coupling strength. Previous studies, which attribute the measured Rabi frequency only to the coherent magnon-photon coupling, may have to be revised. 

Furthermore in the general context, our results suggest that the effect of level attraction, which has been so-far considered as peculiar, might be as ubiquitous as level repulsion. First observed in systems involving inverted oscillators, level attraction has important applications such as topological energy transfer, quantum sensing, and non-reciprocal photon transmission \cite{Glauber1985, Kohler2018, Bernier2017, Fang2017, Barzanjeh2017, Peterson2017, Bernier2018, Xu2018}. Realizing level attraction using solid-state devices has been difficult, but was recently achieved by coupling two opto-mechanical modes to the same dissipative reservoir \cite{Gloppe2014,Xu2016}. It was also proposed \cite{Grigoryan2017} to realize level attraction by engineering the relative phase of microwaves \cite{Wirthmann2010}. Our study shows that level attraction might be generally hidden in systems where coherent coupling dominates self-induced negative feedback (such as the Lenz effect). Engineering and suppressing coherent coupling to reveal level attraction, as demonstrated in our experiment, may pave new ways for creating entangled states, and develop new methods for controlling and utilizing light-matter interactions.       

This work has been funded by NSERC and NSFC (11429401) grants (C.-M.H.). We thank N.R. Bernier, T.J. Kippenberg, P. Barclay, and O. Arcizet for discussions about level attraction in opto-mechanical systems. We also thank K. Xia, V.L. Grigoryan, M. Weides, and I. Boventer for discussions about their independent approach for realizing level attraction by modulating coherent magnon-photon coupling. \\

\end{document}